\documentclass[10pt,letterpaper]{article}
\usepackage{opex3}

\begin{document}

\title{A Monolithic Radiation-Pressure Driven, Low Phase Noise Silicon Nitride Opto-Mechanical Oscillator}

\author{Siddharth Tallur, Suresh Sridaran and Sunil A. Bhave}

\address{OxideMEMS Laboratory, School of Electrical and Computer Engineering, Cornell University, \\ Ithaca, NY, 14853}

\email{sgt28@cornell.edu} 



\begin{abstract}
Cavity opto-mechanics enabled radiation pressure (RP) driven oscillators shown in the past offer an all optical Radio Frequency (RF) source without the need for external electrical feedback. However these oscillators require external tapered fiber or prism coupling and non-standard fabrication processes. In this work, we present a CMOS compatible fabrication process to design high optical quality factor opto-mechanical resonators in silicon nitride. The ring resonators designed in this process demonstrate low phase noise RP driven oscillations. Using integrated grating couplers and waveguide to couple light to the micro-resonator eliminates 1/f$^3$ and other higher order phase noise slopes at close-to-carrier frequencies present in previous demonstrations. We present an RP driven OMO operating at 41.97MHz with a signal power of -11dBm and phase noise of -85dBc/Hz at 1kHz offset with only 1/f$^2$ noise down to 10Hz offset from carrier.
\end{abstract}



\section{Introduction}

Mechanical oscillators coupled to the electromagnetic mode of a cavity have emerged as an important new frontier in photonics, and have enabled interesting experiments in cavity opto-mechanics \cite{kippvahal}. A notable contribution that has come out of this area of research is the manifestation of parametric instability in such resonators, resulting in mechanical amplification and thereby oscillation of the mechanical mode driven purely by pressure of light.

Light can excite oscillations in mechanical resonators in several ways - radiation pressure (RP) driven \cite{mani06}, via gradient forces \cite{kipp09, weig09} or through electrostriction \cite{matt09} and other nonlinear opto-mechanical interactions \cite{sawomo11}. RP driven oscillations have been extensively studied \cite{kippvahal} and such oscillations have been observed by various teams \cite{mani06, eichen09}. All these demonstrations employ either a prism coupler \cite{sawomo11} or a tapered fiber \cite{mani06, kipp09, matt09} to couple light into the microresonator, and are not integrated solutions. Chip-scale opto-mechanical resonators in silicon nitride with integrated waveguides have been shown \cite{gustavo09}, however these devices employ inverse fiber taper coupling and no self-oscillations have been reported. Some solutions also require a CO$_2$ laser reflow step \cite{mani06} to microfabricate the resonator, which is a serial process step and is thus undesirable for a high-yield solution. The phase noise numbers reported for RP driven oscillations \cite{mani06} are far worse than those reported for CMOS oscillators in similar frequency ranges. The phase noise also shows higher order slopes at close to carrier frequencies, that are unexplained but are often attributed to environmental noise. Theoretical analysis for OMOs \cite{mani06, tallur10} indicates that these problems can be overcome by having a high optical quality factor, a moderately high mechanical quality factor, large power handling capacity, and a robust device that is immune to environmental disturbances. As such, there is a need for a truly integrated, monolithic device that can deliver all of these requirements and be compatible with existing chip-scale laser technology to enable a truly chip-scale opto-mechanical oscillator. Keeping all these requirements in mind, we chose Si$_{3}$N$_{4}$ as the material to micro-fabricate the opto-mechanical resonator, as it has been shown to demonstrate high optical \cite{lipson09} and high mechanical \cite{craighead06} quality factors. We present a fabrication process to design high optical quality factor (Q) released opto-mechanical resonators in silicon nitride that demonstrate RP-driven oscillations with low phase noise. Using integrated grating couplers and an integrated waveguide to couple light to the micro-resonator eliminates 1/f$^3$ and other higher order phase noise slopes for close-to-carrier frequencies. Also, our fabrication process flow does not incorporate the CO$_2$ laser reflow step. This RP driven OMO operates at 41.97MHz with a signal power of -11dBm and phase noise of -85dBc/Hz at 1kHz offset with 1/f$^2$ slope.

\section{Design and fabrication}

\subsection{Device design}

We chose a ring geometry for our resonator owing to its simplicity and the high optical Q it offers \cite{lipson09}. To design for high optical quality factors, we select a wide ring (6$\mu$m) with a large radius (40$\mu$m) and narrow spokes (0.5$\mu$m). This ensures that the Si$_3$N$_4$ ring resonator has a high optical Q as the optical mode is confined inside the ring and experiences low bending loss and scattering loss from the spokes. We perform a finite element method (FEM) simulation in COMSOL to identify the fundamental radial expansion mode of the ring, which can be excited via radiation pressure \cite{mani06}. This mode is simulated to be at 39.8MHz.

\begin{figure}[htbp]
\centering
\includegraphics[width = 2.5in]{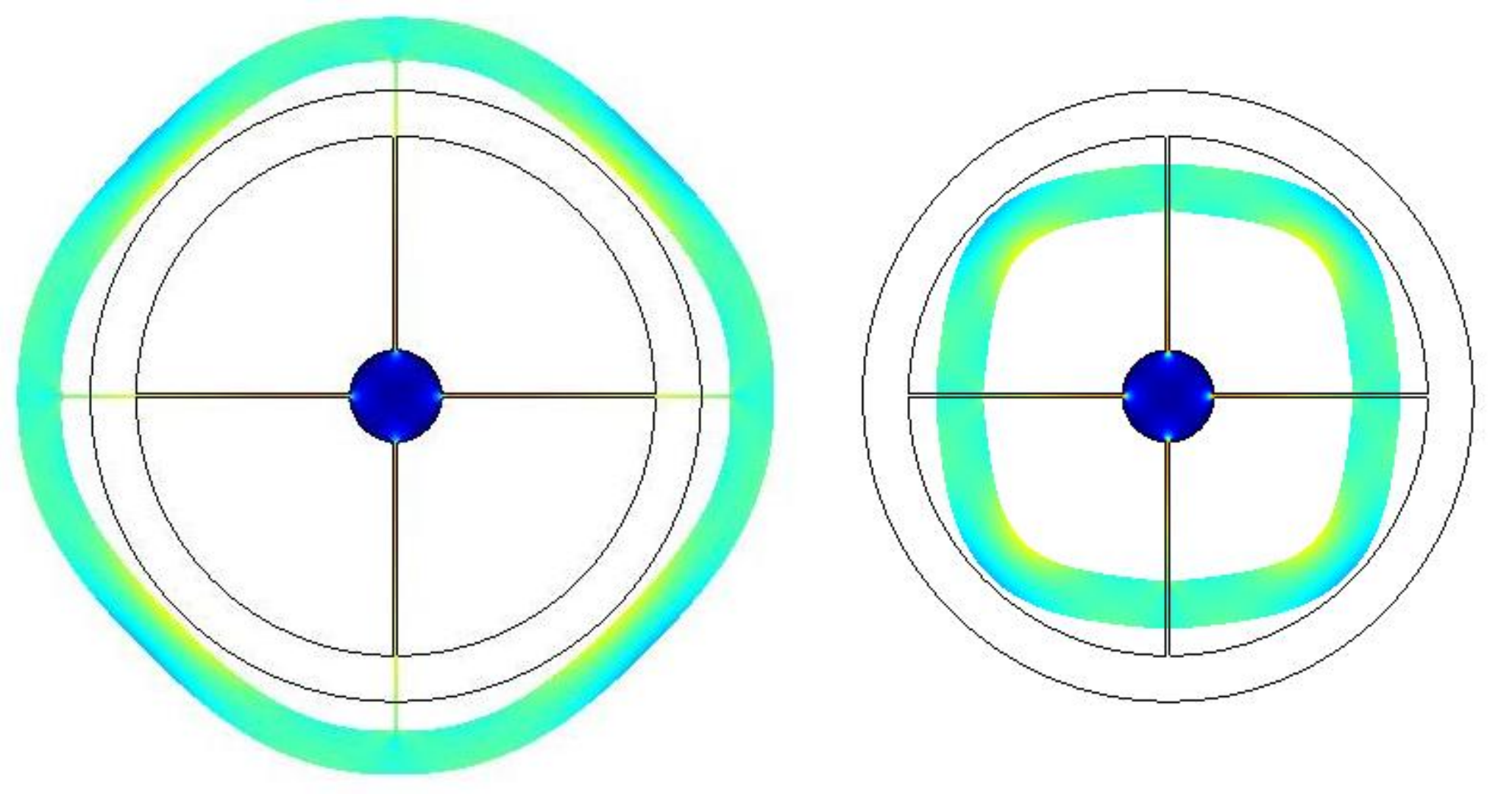}
\caption{Simulated mode shape highlighting deformed geometries for the ring expanding and contracting. The frequency of this fundamental radial expansion mode of the ring is 39.8MHz}
\label{mode}
\end{figure}

The waveguide width at the grating coupler is 15$\mu$m and it is tapered down to 1$\mu$m near the resonator. The grating couplers were designed to be broadband and have maximum transmission around the wavelength of 1550nm, for a fiber angle of 20$^\circ$. We simulated the grating couplers using a commercially available finite difference time domain (FDTD) simulation package (Lumerical FDTD Solutions) and observe insertion loss of -6dB per coupler at 1550nm.

\subsection{Process flow}

\begin{figure}[ht]
\centering
\includegraphics[width = 4in]{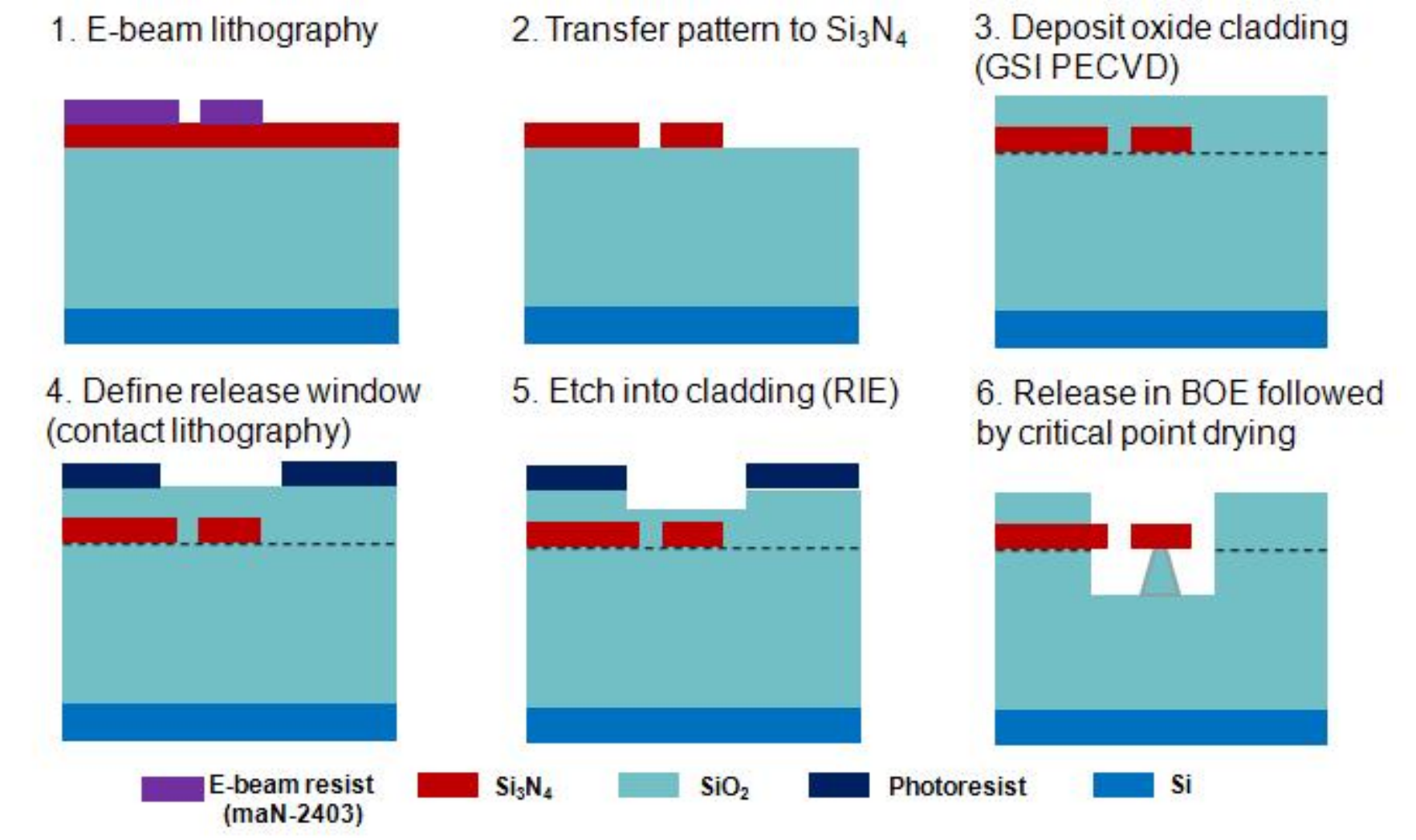}
\caption{Illustration of the process flow to microfabricate grating couplers, waveguides, and opto-mechanical resonators in silicon nitride}
\label{proc_flow}
\end{figure}

We start with silicon wafers that have 4$\mu$m SiO$_2$ thermally grown and deposit 300nm Si$_{3}$N$_{4}$ in an MRL Industries Furnace using low pressure chemical vapor deposition (LPCVD). To define the resonators, waveguides and grating couplers, ma-N 2403 electron beam resist is spun on top of the nitride and patterned using electron beam lithography in a JEOL JBX-9300FS Electron Beam Lithography System. After developing the resist in AZ 726MIF, we etch the pattern into the nitride device layer in an Oxford System 100 Reactive Ion Etcher (RIE) using CHF$_3$/O$_2$ chemistry to etch the nitride. Then we strip the resist and deposit SiO$_2$ cladding using a GSI Plasma Enhanced Chemical Vapor Deposition (PECVD) system, to clad the gratings and waveguide with oxide. This is done to reduce losses at the grating couplers (simulated -10dBm per coupler without cladding; -6dBm per coupler with cladding). A second mask is then used to pattern release windows near the resonator using contact photolithography. This is followed by a partial etch into the cladding in an Oxford PlasmaLab 80+ RIE System using CHF$_3$/O$_2$ chemistry to etch the oxide. This ensures that we can get away with a relatively shorter release time and thereby not have a large undercut for the waveguides. We then perform a timed release etch in buffered oxide etchant to undercut the devices, to enable opto-mechanics. The samples are then dried using a critical point dryer to prevent stiction. Figure \ref{proc_flow} illustrates this process flow schematically.

\begin{figure}[ht]
\begin{minipage}[m]{0.5\linewidth}
\centering
\includegraphics[width = 2.5in]{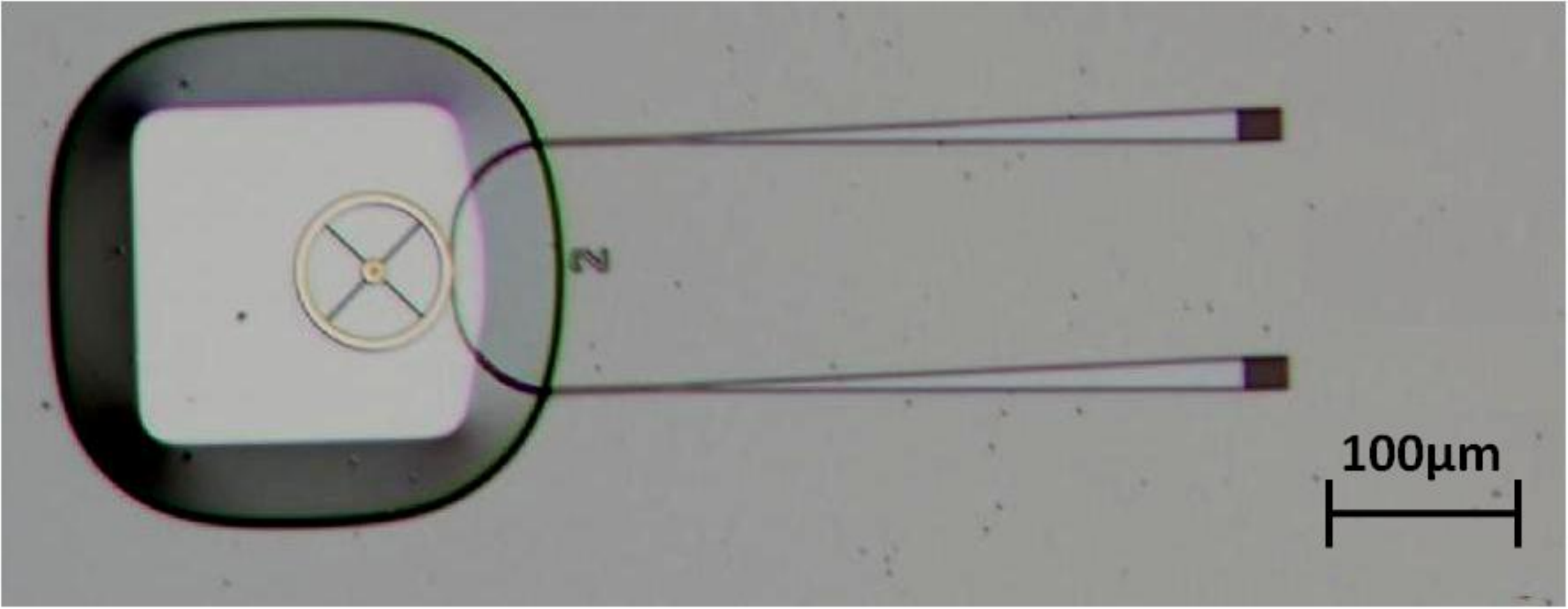}
\end{minipage}
\begin{minipage}[m]{0.5\linewidth}
\centering
\includegraphics[width = 2.5in]{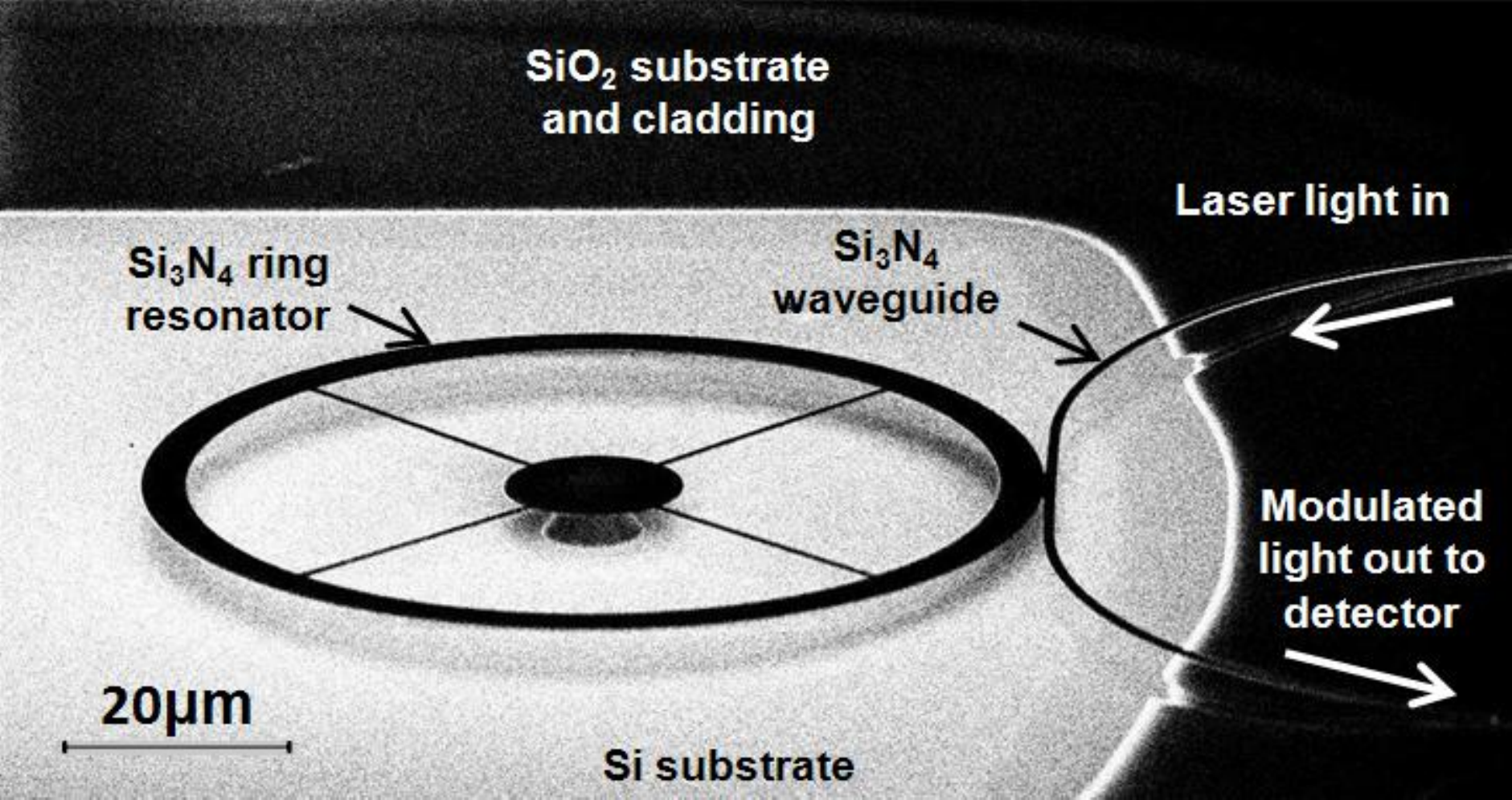}
\end{minipage}
\caption{\textbf{(Left)} Optical micrograph of the integrated device showing the ring resonator, tapered waveguide and grating couplers (top view), \textbf{(Right)} Scanning electron micrograph (SEM) of the ring resonator and released section of the waveguide}
\label{SEM}
\end{figure}

The resulting devices have cladding over the gratings, and the tapered section of the waveguide. The resonator is completely released except for a pedestal in the middle that holds the structure in place. The waveguide is released in the region around the resonator. Figure \ref{SEM} shows an optical micrograph and SEM of the device.

%
%

\section{Experimental setup and measurements}

\subsection{Optical characterization}

\begin{figure}[htbp]
\centering
\includegraphics[width = 5in]{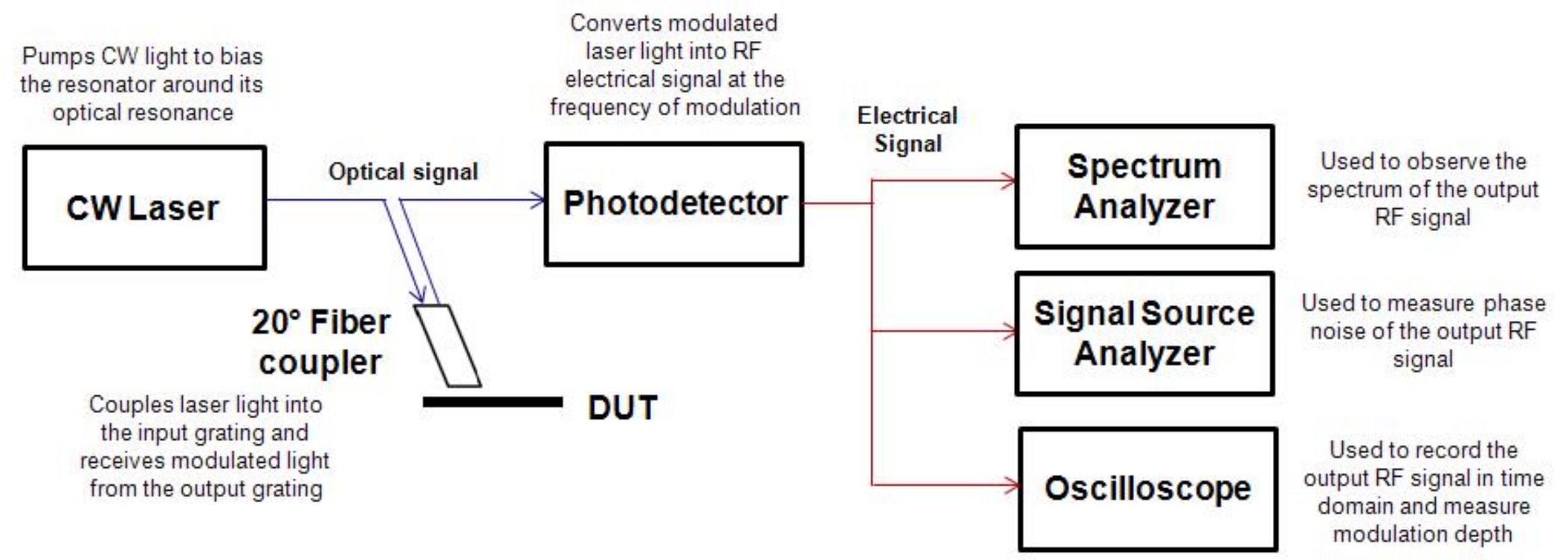}
\caption{Illustration of the test setup to probe the mechanical resonance of the opto-mechanical resonator}
\label{optomech_setup}
\end{figure}

We use the setup of figure \ref{optomech_setup} to study the optical transmission spectrum. We sweep the laser wavelength and observe the output optical power to search for an appropriate high optical quality factor resonance that can be pumped to observe self oscillations. 

\begin{figure}[htbp]
\centering
\includegraphics[width = 2.5in]{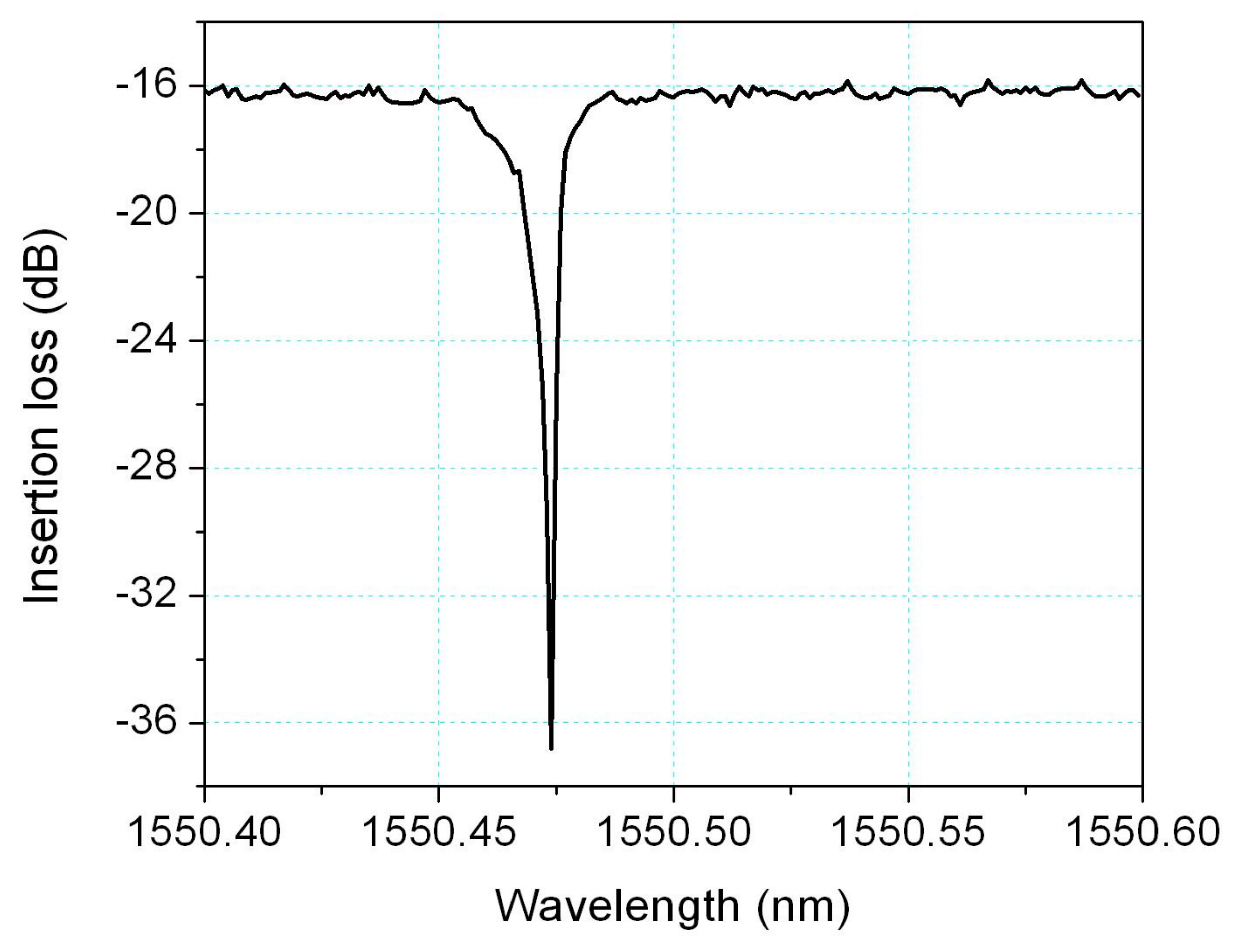}
\caption{Insertion loss measured for -10dBm laser input power for a high loaded Q ($\textgreater$300,000) optical resonance}
\label{optical_sweep}
\end{figure}

Figure \ref{optical_sweep} shows a measurement for insertion loss of a high Q optical resonance around 1550nm measured for -10dBm laser input power. The measured optical loaded Q is $\textgreater$300,000. The off-resonance insertion loss is -8dB per coupler as seen in figure \ref{optical_sweep}. This is due to the refractive index mismatch at the interface where the waveguide transcends from being cladded by oxide to being released, which results in higher losses.

\subsection{Opto-mechanical characterization}

Once we identify a high Q optical resonance to pump, we setup an experiment to measure the mechanical response of this resonator, as illustrated in figure \ref{optomech_setup}. We use an avalanche photodiode (APD) as the photodetector. The laser is biased around the optical resonance of the system to excite the mechanical mode. The gain of the APD is 6,000V/W.

\begin{figure}
\centering
\includegraphics[width = 3in]{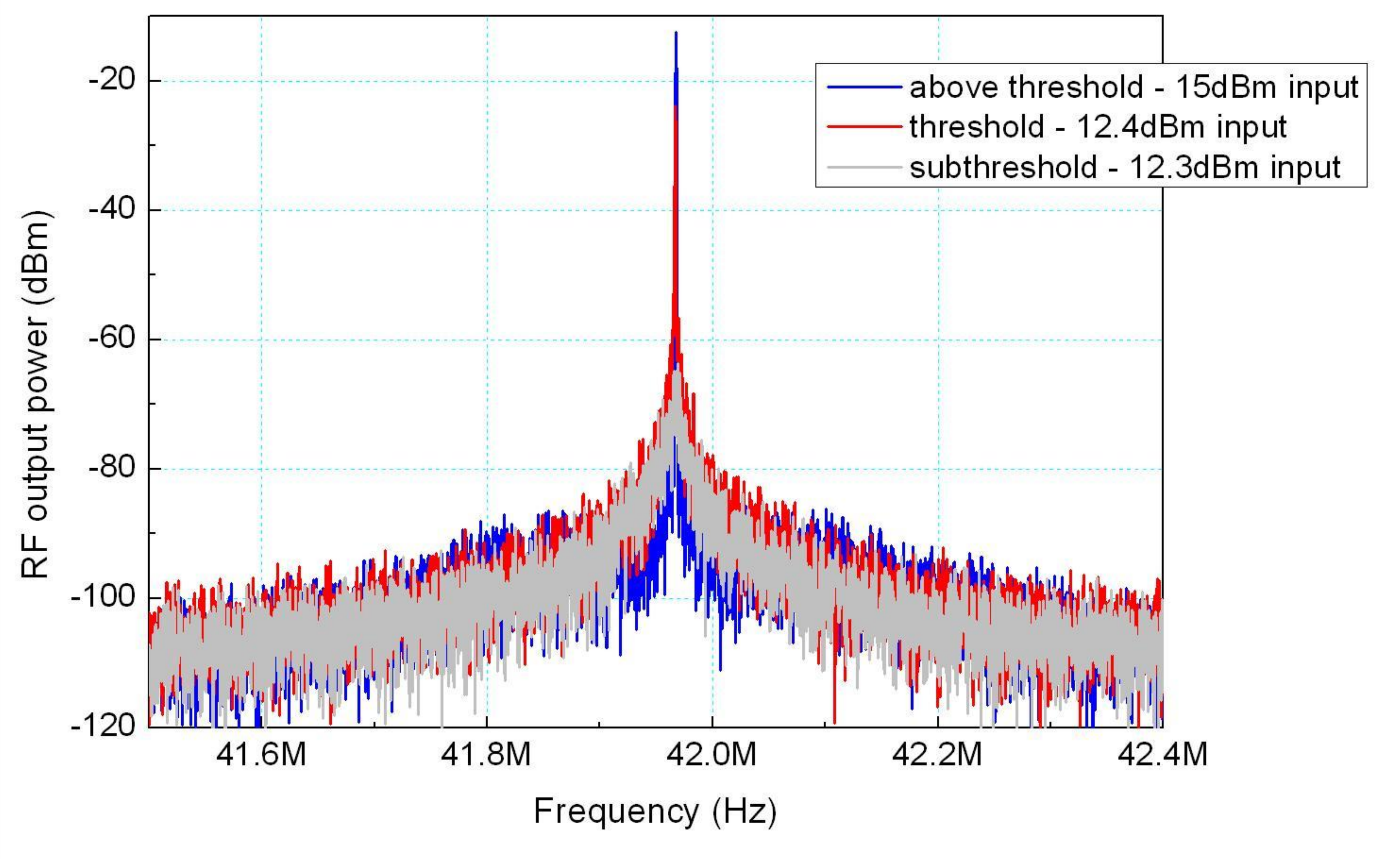}
\caption{Transition from subthreshold brownian noise motion to self oscillation of the mechanical mode of the ring when the laser input exceeds threshold. The laser is biased at a relative detuning $(\Delta\omega/2\delta)$ of 0.38}
\label{combined_SA}
\end{figure}

We set the laser wavelength at a relative detuning of 0.38 from the resonance. Relative detuning is defined as $\Delta\omega/2\delta$. $\Delta\omega$ is the difference between input laser frequency and the optical resonance frequency i.e. $\omega_{in}$-$\omega_{0}$ and $\delta$ is the FWHM linewidth of the optical resonance. We observe that beyond the threshold power, the light coupled into the resonator launches radiation pressure driven self oscillations of the opto-mechanical resonator, as shown in figure \ref{combined_SA}.

\begin{figure}[htbp]
\centering
\includegraphics[width = 3in]{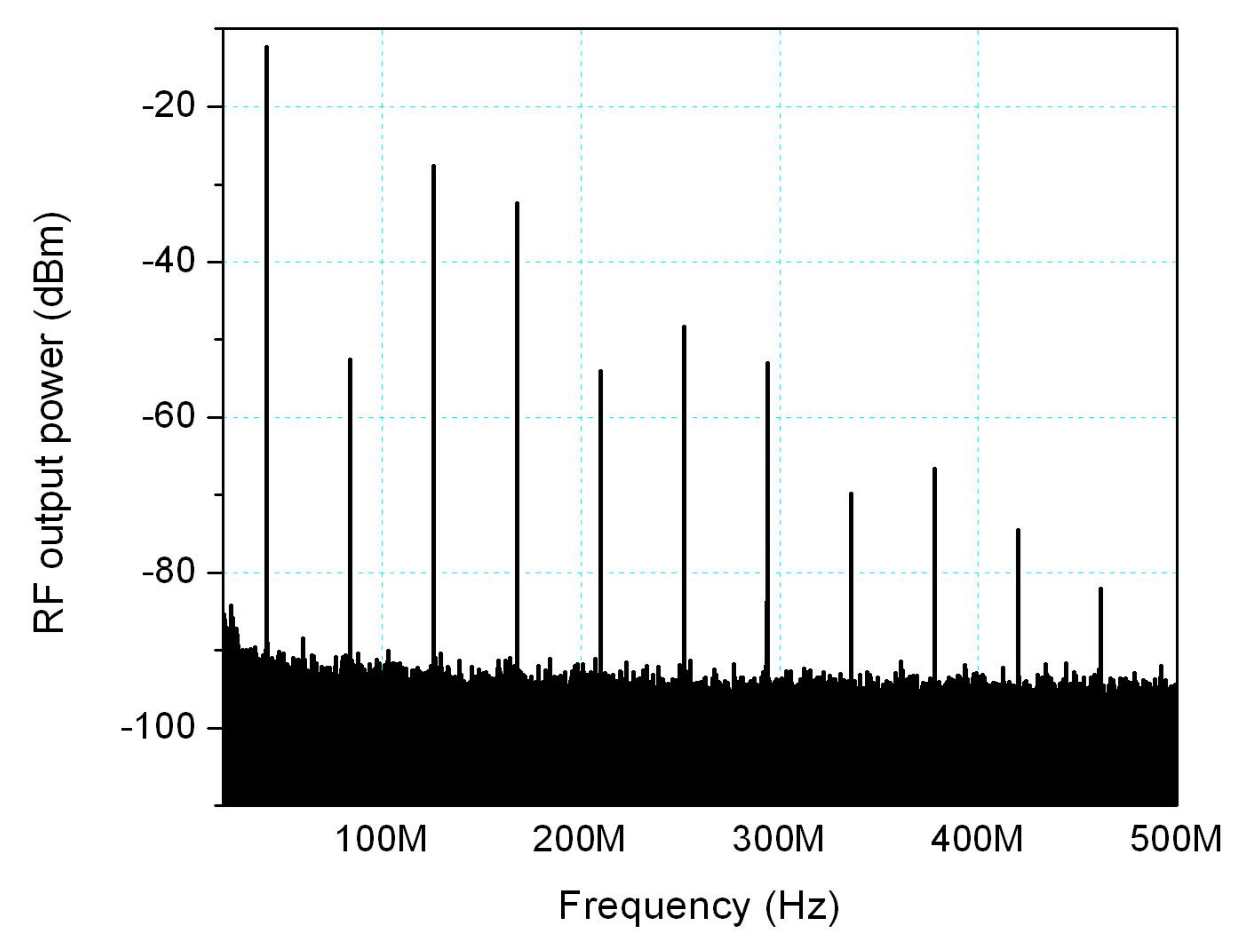}
\caption{Wide frequency sweep showing various harmonics of the fundamental mode for +15dBm laser input power and relative detuning $(\Delta\omega/2\delta)$ of 0.38}
\label{wide_sweep}
\end{figure}

The spectrum is measured on an Agilent E4445A spectrum analyzer. Below the threshold power, the power built up inside the cavity is not enough to cause radiation pressure induced instability, and the amplitude modulation of the CW laser light senses the Brownian noise motion of the micro-ring. The sub-threshold mechanical quality factor was measured to be 2,000. The resonance corresponds to the mode shown in figure \ref{mode}. We perform a wide frequency sweep and see multiple harmonics of the fundamental mode. These harmonics are a result of the non-linear transfer function of the device, which also behaves as an optical modulator \cite{rokhsari06}. The wide frequency sweep is shown in figure \ref{wide_sweep}.

\begin{figure}[htbp]
\centering
\includegraphics[width = 5in]{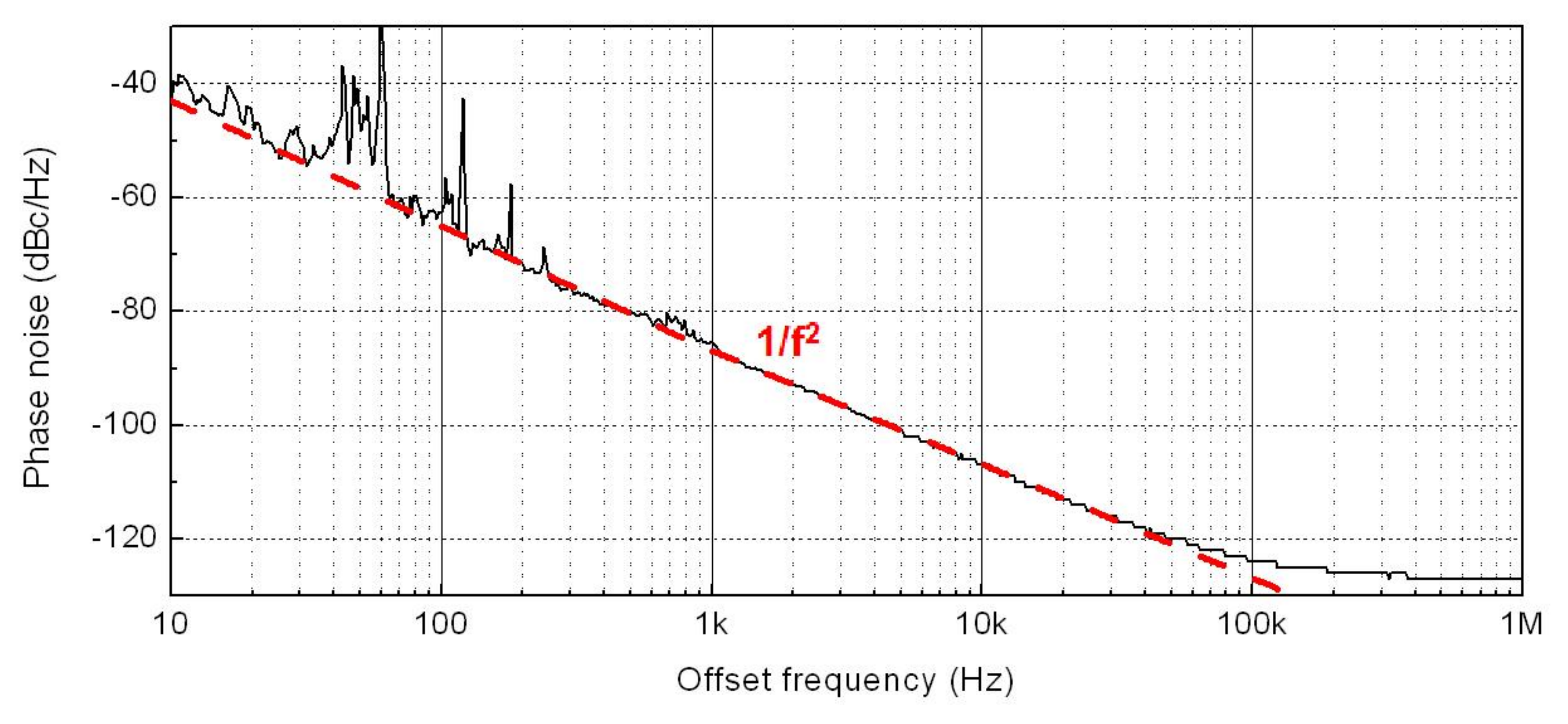}
\caption{Phase noise for OMO operating at 41.947MHz with -11.37dBm output power. The laser input power is +15dBm and it is biased at a relative detuning $(\Delta\omega/2\delta)$ of 0.38. It varies as 1/f$^2$ below 100kHz offsets as indicated by the dotted red trend-line implying that the OMO has no flicker noise. The corner frequency for 1/f$^2$ region is around 20kHz which agrees with the measured mechanical Q of 2,000 at 41.947MHz. The phase noise is measured with an Agilent E5052B signal source analyzer}
\label{phase_noise}
\end{figure}

Figure \ref{phase_noise} shows the phase noise measured for the fundamental mode. The phase noise at 1kHz offset for the 41MHz OMO is -85dBc/Hz. This is better than the phase noise reported for radiation pressure driven self-oscillations of a microtoroid \cite{mani06}. The phase noise varies as 1/f$^2$ below 100kHz offsets as indicated by the dotted red trend-line implying that the OMO has no flicker noise. The corner frequency for 1/f$^2$ region is around 20kHz which agrees with the measured mechanical Q of 2,000 at 41.947MHz. The phase noise is measured with an Agilent E5052B signal source analyzer.

\subsection{Characterization of RP driven oscillations}

\begin{figure}[ht]
\begin{minipage}[m]{0.5\linewidth}
\centering
\includegraphics[width = 2.5in]{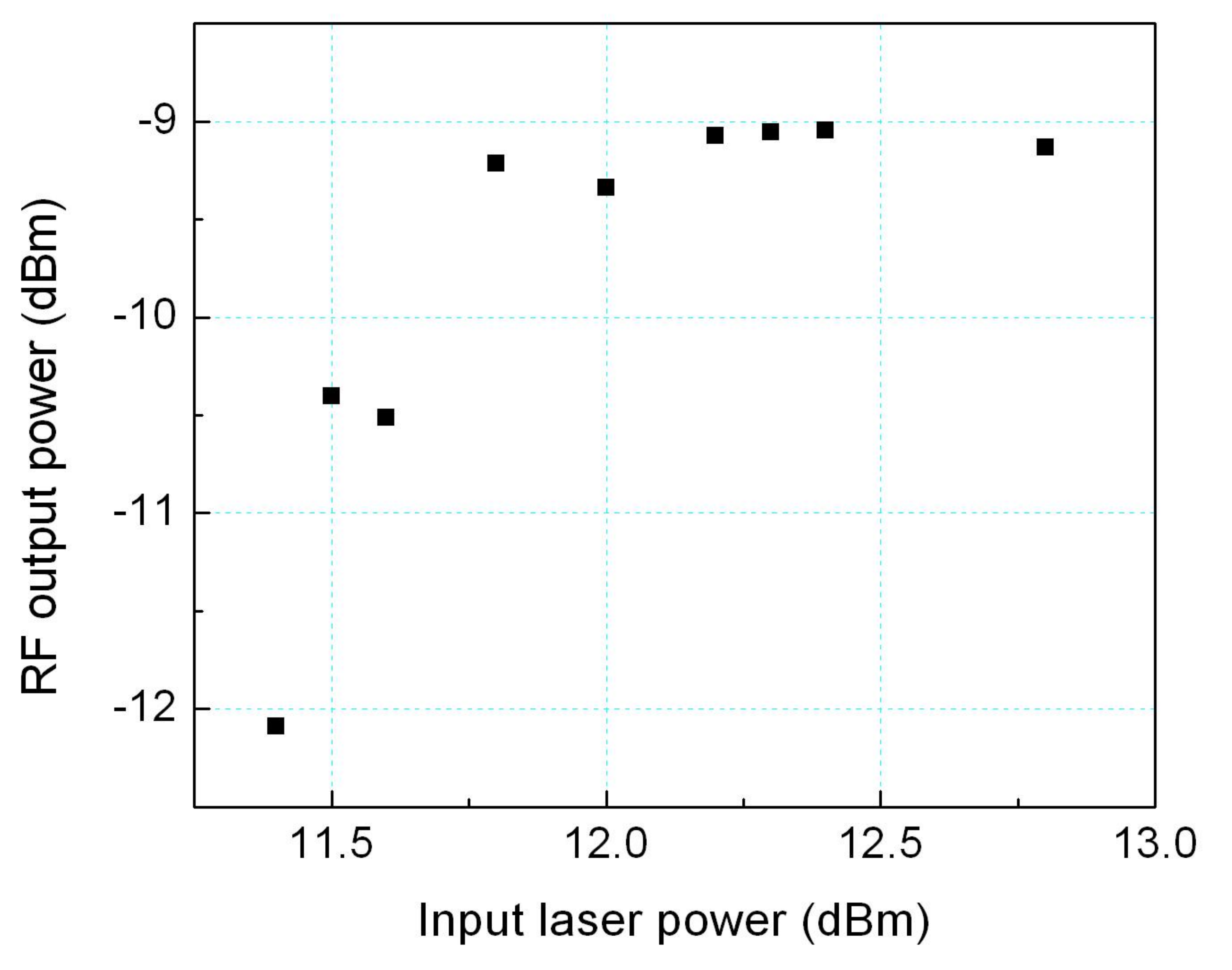}
\end{minipage}
\begin{minipage}[m]{0.5\linewidth}
\centering
\includegraphics[width = 2.5in]{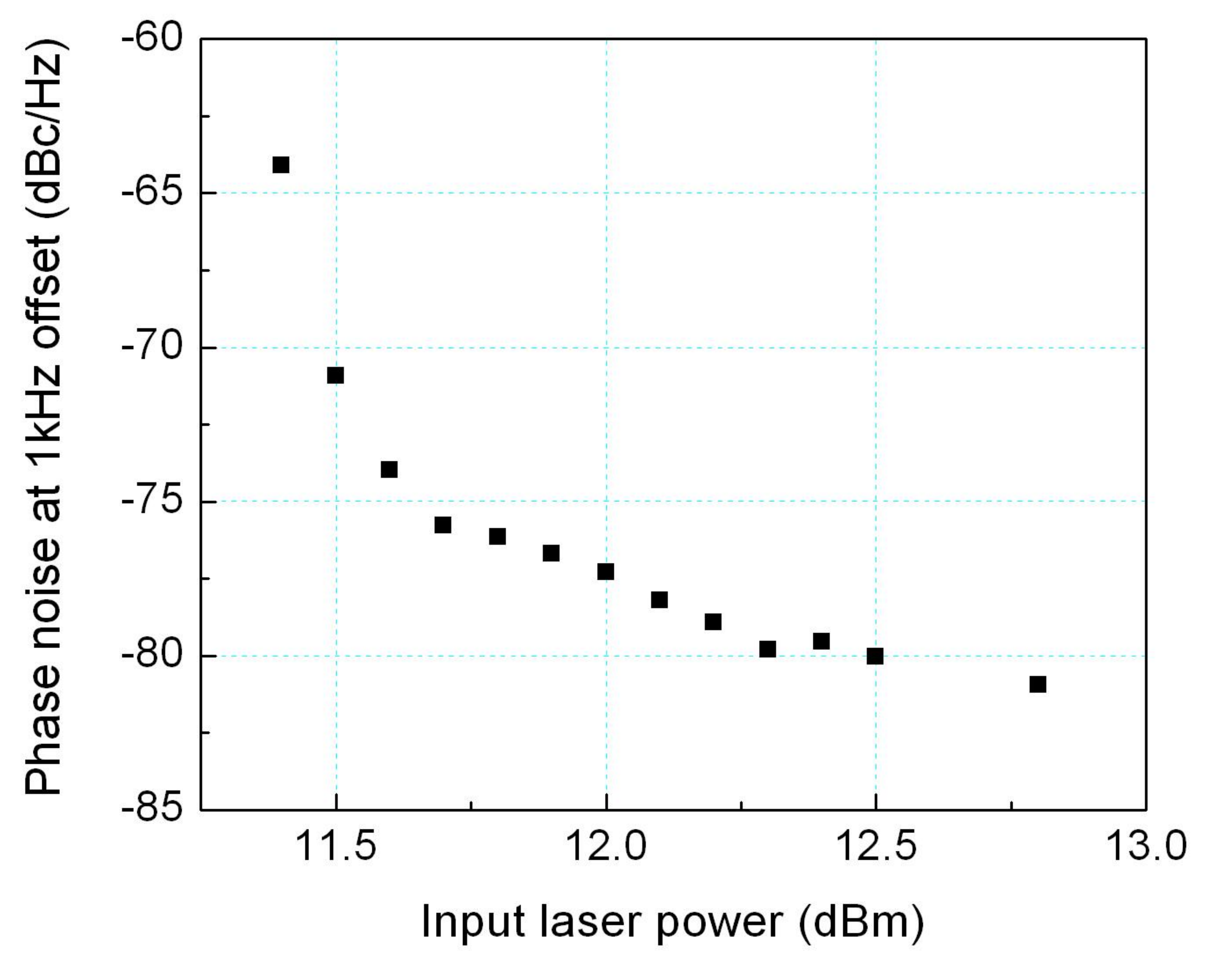}
\label{fig:figure2}
\end{minipage}
\caption{Variation of RF output power \textbf{(Left)} and phase noise at 1kHz offset \textbf{(Right)} with input laser power. The relative detuning was set at 0.31}
\label{plots_ip}
\end{figure}

This section presents characterization experimental data for the RP driven Si$_3$N$_4$ OMO. Figure \ref{plots_ip} show the behavior of RF output power and phase noise as we vary the input laser power, while keeping the relative laser detuning from the optical resonance fixed at 0.31. 

As we see, the RF output power increases as we increase the input laser power and eventually the rate at which it rises tapers off. The phase noise improves as we increase input laser power, eventually following the saturation behavior demonstrated by the RF outputpower. This trend is in good agreement with predictions for radiation pressure driven oscillations in microtoroids \cite{mani06}.

\begin{figure}[htbp]
\centering
\includegraphics[width = 2.5in]{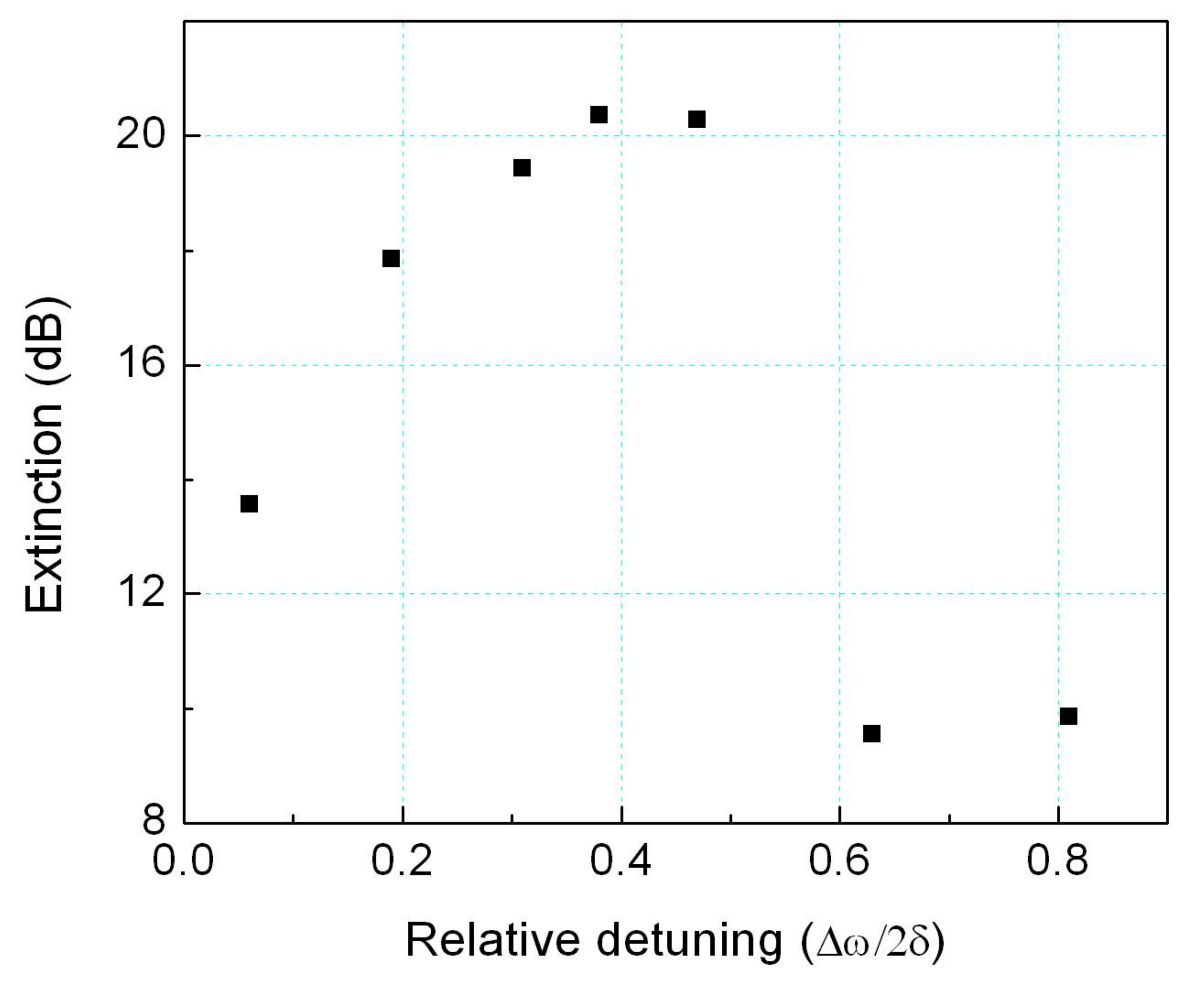}
\caption{Extinction of oscillation waveform at output of the APD measured for various relative detuning values. As we can see, a relative detuning of 0.5 enables us to see maximum extinction of 20dB, which corresponds to extinction at resonance as in figure \ref{optical_sweep}}
\label{mod_depth_detuning}
\end{figure}

Figure \ref{mod_depth_detuning} shows how the extinction of the oscillator varies with relative detuning. Extinction is defined as $E=10log_{10}\frac{V_{max}}{V_{min}}$, where $V_{max}$ and $V_{min}$ are respectively the maximum and minimum voltage levels at the output of the photodetector. As expected \cite{mani06}, we see a maximum extinction of 20dB for relative detuning of 0.5, which corresponds to extinction at resonance as in figure \ref{optical_sweep}. This implies that the radial displacement amplitude of the disk at 0.5 relative detuning is sufficient for complete modulation of the laser light. 

\section{Discussion and Conclusion}

We have presented a low phase noise, integrated monolithic RP driven Si$_3$N$_4$ opto-mechanical oscillator. We designed a process flow to incorporate integrated waveguides and grating couplers along with the micro-ring resonators. The process flow is compatible with typical CMOS fabrication. Integrating the waveguide on-chip also eliminates 1/f$^3$ and other higher order phase noise slopes for close-to-carrier frequencies, that are notorious in demonstrations employing tapered fiber coupling \cite{mani06}. This Si$_{3}$N$_{4}$ micro-ring OMO operates at 41.947MHz with a signal power of -11.37dBm and phase noise of -85dBc/Hz at 1kHz offset with only 1/f$^2$ slope, as compared to quartz oscillators in few MHz range that suffer from high flicker noise \cite{quartz}. The size of the resonator, waveguide and grating coupler system is 600$\mu$m x 150$\mu$m. To the best of our knowledge, this is the first ever observation of radial expansion mode RP driven self-oscillation in silicon nitride micro-rings. The excellent close-to-carrier phase noise performance of this OMO qualifies it as a potential truly on-chip frequency source with high long term stability.

\section{Acknowledgments}

This work was supported by the DARPA/MTO's ORCHID program and Intel. The opto-mechanical device was fabricated at the Cornell NanoScale Science and Technology Facility, which is supported by the National Science Foundation.


\begin{thebibliography}{99}

\bibitem{kippvahal} T. J. Kippenberg and K. J. Vahala, ``Cavity optomechanics: Back-action at the mesoscale,'' Science 29, vol. 321 no. 5893 pp. 1172-1176, August 2008.

\bibitem{cho10} A. Cho, ``Putting light's light touch to work as optics meets mechanics,'' Science 14, vol. 5980 no. 5893 pp. 812-813, May 2010.

\bibitem{mani06} Mani Hossein-Zadeh, Hossein Rokhsari, Ali Hajimiri and Kerry Vahala, ``Characterization of a radiation-pressure-driven micromechanical oscillator,'' Phys. Rev. A 74, 023813 (2006).

\bibitem{kipp09} G. Anetsberger, O. Arcizet, Q. P. Unterreithmeier, R. Rivière, A. Schliesser, E. M. Weig, J. P. Kotthaus and T. J. Kippenberg, ``Near-field cavity optomechanics with nanomechanical oscillators,'' Nature Phys. 5, 909-914 (2009).

\bibitem{weig09} Q. P. Unterreithmeier, E. M. Weig and J. P. Kotthaus, ``Universal transduction scheme for nanomechanical systems based on dielectric forces,'' Nature 458, 1001-1004 (2009).

\bibitem{matt09} M. Tomes and T. Carmon, ``Photonic micro-electromechanical systems vibrating at X-band (11-GHz) rates,'' Phys. Rev. Lett., vol. 102, p. 113601, March 2009.

\bibitem{sawomo11} A. A. Savchenkov, A. B. Matsko, V. S. Ilchenko, D. Seidel and L. Maleki, ``Surface acoustic wave opto-mechanical oscillator and frequency comb generator,'' Opt. Lett. 36, 3338-3340 (2011).

\bibitem{eichen09} M. Eichenfield, J. Chan, R. M. Camacho, K. J. Vahala and O. Painter, ``Optomechanical crystals,'' Nature 462, 78–82 (2009).

\bibitem{gustavo09} G. S. Wiederhecker, L. Chen, A. Gondarenko and M. Lipson, ``Controlling photonic structures using optical forces,'' Nature 462, 633–636 (2009).

\bibitem{tallur10} S. Tallur, S. Sridaran and S. A. Bhave, ``Phase noise modeling of opto-mechanical oscillators,'' IEEE Frequency Control Symposium (FCS 2010), Newport Beach, California, June 2-4, 2010, pp. 268-272.

\bibitem{lipson09} A. Gondarenko, J. S. Levy, and M. Lipson, ``High confinement micron-scale silicon nitride high Q ring resonator,'' Opt. Express 17, 11366-11370 (2009).

\bibitem{craighead06} S. S. Verbridge, J. M. Parpia, R. B. Reichenbach, L. M. Bellan and H. G. Craighead, ``High quality factor resonance at room temperature with nanostrings under high tensile stress,'' J. Appl. Phys. 99, 124304 (2006).

\bibitem{rokhsari06} H. Rokhsari, T. J. Kippenberg, T. Carmon and K. J. Vahala, ``Theoretical and experimental study of radiation pressure-induced mechanical oscillations (parametric instability) in optical microcavities,'' IEEE J. Sel. Top. Quantum Electron. 12, 96-107 (2006).

\bibitem{quartz} Low Phase Noise Quartz Crystal Oscillator, Model FE-102A. http://www.freqelec.com/qz\_osc\_fe102a.html

\end{thebibliography}
\end{document}